\documentclass[reprint,amsmath,amssymb,aps,floatfix,superscriptaddress,prl, nofootinbib]{revtex4-2}
\usepackage[utf8]{inputenc}
\usepackage[T1]{fontenc}
\usepackage{microtype}
\usepackage{physics}
\usepackage{lmodern}
\usepackage{float}
\usepackage{graphicx}
\usepackage{textcomp}
\usepackage[hidelinks, bookmarks=false]{hyperref}
\usepackage[capitalise]{cleveref}
\usepackage{siunitx}

\begin{document}
\title{Model-based Optimization of Superconducting Qubit Readout}
\author{Andreas Bengtsson}
\author{Alex Opremcak}
\author{Mostafa Khezri}
\author{Daniel Sank}
\author{Alexandre Bourassa}
\author{Kevin J. Satzinger}
\author{Sabrina Hong}
\author{Catherine Erickson}
\author{Brian J. Lester}
\author{Kevin C. Miao}
\affiliation{Google Quantum AI, Santa Barbara, CA}

\author{Alexander N. Korotkov}
\affiliation{Google Quantum AI, Santa Barbara, CA}
\affiliation{Department of Electrical and Computer Engineering, University of California, Riverside, CA}

\author{Julian Kelly}
\author{Zijun Chen}
\author{Paul V. Klimov}
\affiliation{Google Quantum AI, Santa Barbara, CA}

\date{\today}

\begin{abstract}
Measurement is an essential component of quantum algorithms, and for superconducting qubits it is often the most error prone. Here, we demonstrate model-based readout optimization achieving low measurement errors while avoiding detrimental side-effects. For simultaneous and mid-circuit measurements across 17 qubits, we observe 1.5\% error per qubit with a \SI{500}{ns} end-to-end duration and minimal excess reset error from residual resonator photons. We also suppress measurement-induced state transitions achieving a leakage rate limited by natural heating. This technique can scale to hundreds of qubits and be used to enhance the performance of error-correcting codes and near-term applications.

\end{abstract}

\maketitle

Superconducting qubits have achieved measurement errors below 1\% for single qubits \cite{walter2017rapid, sunada2022fast, chen2023transmon, swiadek2023enhancing} thanks to advancements including dispersive readout \cite{blais2021circuit} and quantum-limited parametric amplifiers \cite{aumentado2020superconducting}. However, the increasing scale and complexity of algorithms bring a new set of challenges beyond simple single-qubit measurements. For example, quantum error correction requires measurements to be performed simultaneously, in the middle of circuits, and in a repetitive fashion. 
Mid-circuit measurements must be fast to avoid decohering qubits not being measured, but faster readout can increase measurement and leakage errors. In recent error-correction experiments with superconducting qubits, readout-induced leakage severely limited performance \cite{sundaresan2022matching, marques2023all}, and in another, 20\% of the total error (twice that of the measurement itself) was due to qubit idling during measurement \cite{google2023suppressing}.

Typically, readout is calibrated in-situ, meaning that control parameters like pulse amplitude and frequency are varied until the observed measurement error is minimized. While effective at minimizing measurement error for isolated qubits, it can fail to capture other destructive processes like leakage or residual resonator photons. Additionally, effects such as qubit-qubit coupling impose non-locality in that the optimal values for one qubit depend on the values of neighboring qubits. In turn, qubits optimized in isolation tend to perform poorly when measured simultaneously. Thus, optimizing multi-qubit measurement requires searching a parameter space where the dimension scales linearly with the number of qubits, while attempting to minimize many metrics at once. This task rapidly becomes intractable by in-situ parameter sweeps as the number of qubits grows.

In this Letter, we present an ex-situ (model based) optimization technique for readout parameters. Ex-situ optimization which allows us to explore a larger parameter space and minimize errors that are difficult or costly to measure, compared to in-situ optimization where the speed is limited by the data rate of the quantum processor.
In designing our model-based approach, we tackle three challenges which are often seen in quantum optimal control \cite{koch2022quantum}. 
First, if the models do not accurately capture the present error channels, ex-situ optimization is likely to perform worse than in-situ. 
Second, the models must be evaluated quickly to be able to actually explore a larger space. These two challenges are conflicting in that more accurate models typically result in slower evaluation; for instance, a quantum simulation of the system dynamics would be too slow. 
Third, we must use an optimization algorithm that can find a reasonably good minimum without requiring a prohibitively long runtime.

We begin by describing representative models for readout error channels relevant to a Sycamore processor \cite{arute2019quantum}, which consists of superconducting frequency-tunable transmon qubits, each coupled to their own readout resonator. The models are quick to evaluate and we demonstrate that they accurately predict a variety of metrics over a wide range of parameters. We then use them together with the snake optimizer \cite{klimov2020snake} to minimize the errors for 17 qubits in a distance-3 surface code. We achieve 1.5\% measurement error per qubit in \SI{500}{ns} (from the start of the readout until the system is ready for the next operation), while also reducing any additional errors like reset and leakage. 

\begin{figure}
    \centering
    \includegraphics[width=\linewidth]{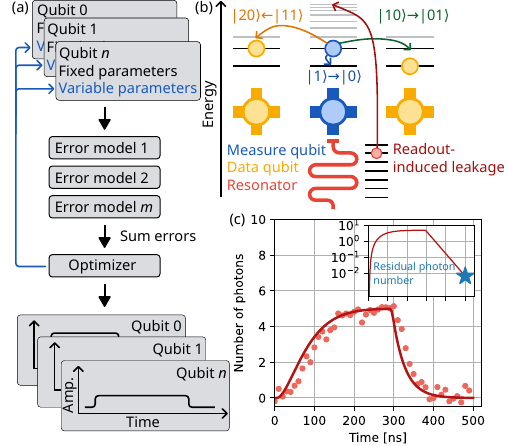}
    \caption{
    (a) The optimization workflow. We build error models from fixed parameters like resonator frequency and linewidth, which we then run an optimizer on to find a set of variable parameters (e.g. pulse amplitude and length) that gives low errors. The output of the optimizer is a unique pulse shape for each qubit, as well as qubit frequency (not shown).
    (b) Examples of readout error mechanisms. The middle qubit, which is in $|1\rangle$, is coupled to two other qubits. During readout of the middle qubit its excitation can relax to $|0\rangle$; swap into $|1\rangle$ of the right qubit; or combine with the excitation in the left qubit to $|2\rangle$. Additionally, photons in the readout resonator can excite the qubit up to a high state.
    (c) Number of photons in the readout resonator as a function of time, both simulated (solid line) and measured (circles). The inset shows the simulated values on a logarithmic scale. Residual photons can cause reset and dephasing errors.
    }
    \label{fig:mechanisms}
\end{figure}

The models fall into two categories: predictive or heuristic. Ideally, we would only have predictive models (models that accurately predict error rates), but this is not always feasible as the computation might be too inaccurate or take too long. In those cases we use heuristic models to steer the optimizer away from parameter regions where errors are large, but difficult to quantify. We use predictive models for the signal-to-noise ratio (SNR), qubit relaxation during readout, and residual resonator photons.  Heuristic models include measurement-induced state transitions \cite{shillito2022dynamics, khezri2023measurement}, and coupling to neighboring qubits. We sum all models into a single optimization cost function. The process is illustrated in \cref{fig:mechanisms} (a).

Each model takes one or several input parameters describing the properties of the qubits, their readout resonators, or the measurement system itself. In total there are seven such parameters,
\begin{itemize}
    \item Qubit anharmonicity, $\alpha < 0$
    \item Resonator-qubit coupling, $g(\omega_q)$
    \item Bare resonator frequency, $\omega_r$
    \item Measurement efficiency, $\eta$ \cite{blais2021circuit}
    \item Resonator linewidth, $\kappa$
    \item Qubit relaxation rate as a function of frequency, $\Gamma_1(\omega_q)$
    \item A calibrated reference for the readout pulse power at the processor
\end{itemize}
We characterize these using a suite of metrology experiments \cite{sank2024characterization}. Most of them are static and characterized just once. However, the qubit relaxation time is known to fluctuate \cite{burnett2019decoherence} and is therefore remeasured just before optimization.

Additionally, there are four parameters which we can tune and optimize,
\begin{itemize}
    \item Qubit frequency during readout, $\omega_q$
    \item Readout pulse amplitude, $B_0$
    \item Readout pulse length, $t_p$
    \item Readout ringdown length, $t_r$
\end{itemize}
The ringdown length is needed for mid-circuit measurements to allow the resonator to decay back to its ground state before other operations can resume. We choose to use a fixed total readout time ($t_p + t_r= \SI{500}{ns}$), allowing us to synchronize gates, as well as reduce the number of optimization parameters to three.

A key parameter derived from the model inputs is the separation between resonator frequencies for the states $|0\rangle$ and $|1\rangle$, i.e. the dispersive shift $2\chi(\omega_q)$, given by \cite{khezri2018dispersive},
\begin{equation}
    \label{eq:chi}
    \chi(\omega_q) =  \frac{g(\omega_q)^2 \alpha}  {(\omega_q - \omega_r)^2  (1 + \alpha / (\omega_q - \omega_r))}  \left(1 - \frac{\omega_q - \omega_r}{\omega_q}\right).
\end{equation}
The shift is tuneable since it depends on the qubit frequency; absent other constraints, the SNR per measurement photon is maximized when $2\chi=\kappa$.

A second derived parameter is the field in the resonator as a function of time, $\beta(t)$ which is found by solving
\begin{equation}
    \label{eq:diff_eq}
    \frac{d \beta}{d t} = \sqrt{\kappa} B(t) + (i\Delta - \kappa / 2) \beta(t),
\end{equation}
where $B(t)$ is the readout drive amplitude and has the dimension $\sqrt{\mathrm{photons~per~time}}$, and $\Delta$ is the frequency difference between the drive and the dressed resonator. In the rest of this paper we restrict the drive to be in the center of the resonator frequencies corresponding to $|0\rangle$ and $|1\rangle$, i.e. $\Delta=\pm \chi$, since that yields the highest SNR in the parameter regime we are interested in. For both states, we find the corresponding $\beta_{|0\rangle}(t)$ and $\beta_{|1\rangle}(t)$ by numerically solving \cref{eq:diff_eq}.

The applied readout pulse, together with the noise (assumed to be Gaussian) in the readout chain, leads to a certain SNR, from which we derive the corresponding probability to misidentify the state. Given ${\delta \beta(t) = \beta_{|0\rangle}(t) - \beta_{|1\rangle}(t)}$, the SNR is calculated as
\begin{equation}
    \text{SNR} = 2 \eta\kappa \frac{ \left| \int_0^{t_p + t_r} \delta \beta(t)w(t) dt \right| ^2 }{\int_0^{t_p + t_r}|w(t)|^2},
\end{equation}
where $w(t)$ is an integration window function, which we set to $\delta\beta(t)^*$ to maximize $\text{SNR}$. From $\text{SNR}$ we derive the corresponding error,
\begin{equation}
    \label{eq:snr}
    \epsilon_{\text{separation}} = \frac{1}{2} \text{erfc}\left( \frac{\sqrt{\text{SNR}}}{2} \right).
\end{equation}

During readout the qubit might decay and potentially cause a measurement error. To calculate the error rate we need the qubit frequency during readout, which is changing throughout the process due to the AC-Stark effect, and the corresponding relaxation rates at those frequencies. The latter is measured by a standard relaxation experiment versus qubit frequency; while the former can be found via \cref{eq:chi,eq:diff_eq},
\begin{equation}
    \label{eq:freq}
    \omega_q(t) = \omega_q(0) + 2 |\beta_{|1\rangle}(t)|^2 \chi(\omega_q(0)),
\end{equation}
where we have assumed that the AC-Stark shift is strictly linear.
Given $\Gamma_1(\omega_q)$, we calculate the relaxation error as
\begin{equation}
    \label{eq:relax}
    \epsilon_{\text{relaxation}} = \int_0^{T_0} \Gamma_1(\omega_q(t)) \mathrm{d}t.
\end{equation}
We approximate $T_0$ to be the point where $\text{SNR}$ is half of its maximum, since a relaxation event beyond that point should not change the measurement outcome. We also choose to ignore the upward transition rate $|0\rangle \rightarrow|1\rangle$, since the timescale for that process is much longer than for relaxation and the readout.

The resonator must be mostly depleted of photons before the next operation can begin, since any remaining photons cause qubit dephasing. Additionally, in our architecture such photons directly translate into reset errors since it is based on the swap interaction between the qubit and its resonator \cite{mcewen2021removing}, and a photon in the resonator could be swapped into either $|1\rangle$ or $|2\rangle$, depending on the state. We use the mean photon number in the resonator at the end of the readout,
\begin{equation}
  \epsilon_{\text{photon}} = \frac{|\beta_{|0\rangle}(T)|^2 + |\beta_{|1\rangle}(T)|^2}{2},
\end{equation}
as the error model to minimize both the reset error and qubit dephasing.

Shown in \cref{fig:mechanisms}(c) are the expected photon number, $|\beta_{|0\rangle}(t)|^2$, and the corresponding measured values (via spectroscopic measurements of the AC-Stark shift \cite{khezri2023measurement}). The measurement technique is not sensitive to the small frequency shifts occurring at the low photon numbers towards the end of readout; however, since the agreement is good during the pulse itself we can use the model to infer what the final photon number is, which in this example is 0.005.

\begin{figure}
    \centering
    \includegraphics[width=\linewidth]{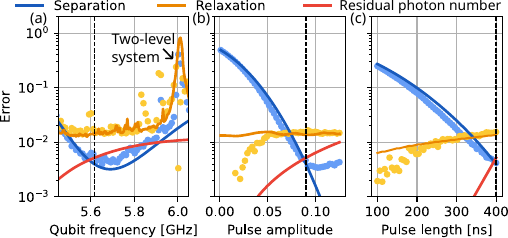}
    \caption{Error models and their dependence on readout parameters.
    (a) Separation and relaxation errors, and residual photon number, versus the qubit frequency with amplitude and length kept fixed. Circles are measured data, and lines are simulated values. As the qubit frequency changes, we track the readout pulse frequency to be centered between the two dressed resonator states. The peak at \SI{6}{GHz} is due to a two-level system defect.
    (b) and (c) show the same models, but versus pulse amplitude and length, respectively. The total readout time is kept constant, such that when the pulse length increases, the ringdown time decreases. The non-swept parameters are kept fixed at the values indicated by the dashed vertical lines in the respective panels.
    }
    \label{fig:models}
\end{figure}

In \cref{fig:models}, we show predicted and measured values for $\epsilon_\text{separation}$ and $\epsilon_\text{relaxation}$ as a function of qubit frequency, pulse amplitude, and pulse length. We additionally show the predicted residual photon number, though we do not have a sensitive enough technique to reliably measure this quantity at the modeled levels.
Overall, we see good agreement between measured and simulated values, with the exception for $\epsilon_\text{relaxation}$ at low amplitudes and lengths where $\epsilon_\text{separation}$ is large. That parameter regime should be avoided and accurately predicting $\epsilon_\text{relaxation}$ there is less important. The cause of the discrepancy could be due to the approximations in \cref{eq:relax} or from the inaccuracy in trying to extract a small error on top of a large error.

We can understand the tradeoffs in readout optimization by studying the predictive models. If we only consider these three models, the ideal pulse would be short and with high amplitude, since the SNR approximately scales quadratically with the amplitude [\cref{fig:models} (b)] and linearly with the pulse length [\cref{fig:models} (c)], while $\epsilon_{\text{photon}}$ also scales quadratically with the amplitude, but exponentially with the pulse length (for a fixed total time). Additionally, a short readout pulse minimizes $\epsilon_{\text{relaxation}}$.

However, a high pulse amplitude can be problematic for several reasons. For example, it leads to measurement-induced state transitions \cite{shillito2022dynamics, khezri2023measurement}, which occurs when resonator photons are transferred to the qubit and excite it far beyond the computational subspace, as illustrated in \cref{fig:mechanisms} (b). While this high state may lead to a measurement error, it is more importantly immune to our reset protocol \cite{mcewen2021removing}, making it particularly destructive for mid-circuit measurements and quantum error correction \cite{sundaresan2022matching, marques2023all}. Using the model in Ref.~\cite{khezri2023measurement}, valid only for $\omega_q > \omega_r$, we define a heuristic that constrains the maximum photon number in the resonator,
\begin{equation}
   \text{max}(|\beta(t)|^2) < a e ^ {b (\omega_q - \omega_r)} - \sqrt{a e ^ {b (\omega_q - \omega_r)}},
\end{equation}
where $a$ and $b$ are extracted from numerical simulations and only dependent on $\alpha$ and $g$ \cite{khezri2023measurement}.

Finally, we introduce a model related to the coupling between qubits. Our qubits are laid out on a square grid where a pair of qubits have four relevant coupling channels, $|01\rangle \leftrightarrow |10\rangle$, $|11\rangle \leftrightarrow |20\rangle$, $|11\rangle \leftrightarrow |02\rangle$, and $|12\rangle \leftrightarrow |21\rangle$. 
We heuristically model the errors associated with these channels using a sum of Lorentizians,
\begin{equation}
    \epsilon_{\text{coupling}} = \sum_{i} c_i \frac{\gamma_i}{2} \frac{\pi}{(\omega_q - \omega_i)^2 + \gamma_i^2/4},
\end{equation}
where $c_i$, $\gamma_i$ and $\omega_i$ are the amplitude, width, and center frequency of each transition. 
We use a heuristic to avoid having to model the time dependence of the qubit frequency and its effect on the measurement errors. That dependence is complicated by the AC-Stark effect, which imposes both a frequency shift due to mean number of photons in the resonator [\cref{eq:freq}], as well as frequency broadening due to photon number fluctuations. By using wide and large enough Lorentzians the optimizer avoids any qubit-qubit interactions. Assuming couplings between nearest and next-nearest neighbors there are up to 32 frequency collisions for each qubit.

\begin{figure}
    \centering
    \includegraphics[width=\linewidth]{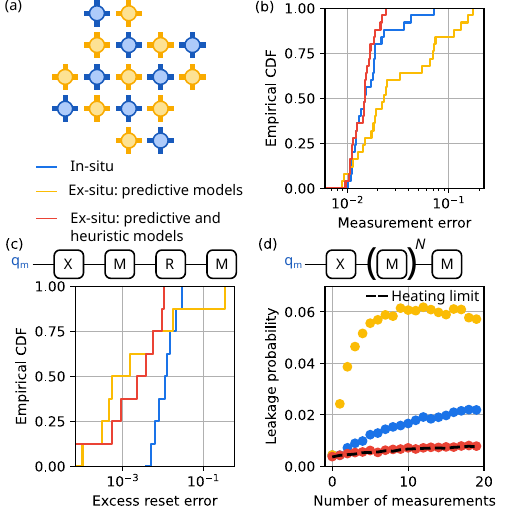}
    \caption{Benchmarking of the optimized readout performance. We compare three optimization strategies: in-situ; ex-situ with only predictive models; ex-situ with predictive and heuristic models.
    (a) The distance-3 surface code with 9 data qubits (yellow) and 8 measure qubits (blue), used for the benchmarking.
    (b) Simultaneous measurement errors for two cases: all qubits, only measure qubits. We prepare a set of random states across the qubits and perform simultaneous measurements. The data shows the combination of the two cases.
    (c) Reset error added by a preceding measurement, benchmarked on the measure qubits. The excess reset error is caused by residual photons in the readout resonator. 
    (d) Average leakage probability in the measure qubits after preparing $|1\rangle$ and performing $N$ measurements. The dashed line shows the heating limit where the measurements are replaced by an equivalent amount of waiting time.
    }
    \label{fig:results}
\end{figure}

We now continue to the actual optimization. While any global optimizer can be used, we choose to employ the snake optimizer \cite{klimov2020snake}, which has successfully optimized single and two-qubit gate parameters for a variety of quantum algorithms \cite{arute2019quantum, google2023suppressing, morvan2023phase}. More optimization details are found in Ref.~\cite{SI}. As our experimental platform we use 17 qubits in a distance-3 surface code layout, illustrated in \cref{fig:results}(a). The optimization takes 1 minute and includes 1.7 million evaluations of the cost function.
Afterwards, the resulting parameters are uploaded to the control system, and the only remaining calibration is to find the discrimination line to distinguish between $|0\rangle$ and $|1\rangle$ for each qubit.
We choose to not model this since we can efficiently measure it simultaneously across all qubits, and it does not conflict with the other parameter choices.

We compare three different optimization strategies to evaluate the performance of our model-based approach. The first strategy is in-situ optimization where we choose a fixed pulse length (\SI{300}{ns}) and perform a sequence of 1D sweeps to find the optimal pulse frequency, amplitude, qubit frequency, and integration window. The second strategy is ex-situ optimization using a partial cost function consisting of only the predictive models, i.e. no qubit-qubit coupling or measurement-induced state transitions models. The third strategy is ex-situ optimization using a complete cost function consisting of all available models. For each strategy, we quantify three important aspects: measurement errors, reset errors, and leakage. Note that we do not benchmark the performance of optimizer itself, e.g. how well it finds the actual global minimum. The performance aspects of the snake have been recently studied in Ref.~\cite{klimov2023snake}. 

We benchmark measurement errors by preparing 200 random initial states over all qubits and then sampling 2,000 measurement outcomes for each initial state. We then repeat the procedure, but this time using only the measure qubits to mimic the surface code mid-circuit measurements.  We compare all outcomes with the known initial states and extract the errors, seen in \cref{fig:results} (b), and calculate the measurement error as $\left(P(1|0) + P(0|1)\right)/2$, where $P(f|i)$ is the probability of preparing $|i\rangle$ and measuring $|f\rangle$. Note that state preparation errors will show up as measurement errors in this protocol. The complete ex-situ optimizer achieves an average measurement error of 1.5\% per qubit, while in-situ and partial ex-situ optimization achieve 1.9\% and 4.7\%, respectively. Overall, in-situ and complete ex-situ optimization have similar performance with the exception of a few high-error outliers for the in-situ optimizer. For instance, the largest outlier is caused by $|11\rangle \leftrightarrow |02\rangle$ swapping between two neighboring qubits, which the ex-situ optimizer is able to avoid \cite{SI}. Out of the 1.5\% error per qubit, we are able to account for 1.2\% when we include the contributions from state preparation, separation error, and relaxation error \cite{SI}. We estimate that the state preparation error is 0.4\%, which, if accurate, should be subtracted from the values above. However, since state preparation errors should affect all three strategies the same we have chosen to be conservative and not do the subtraction.

Next, we benchmark reset errors added by readout for the measure qubits only (data qubits do not need reset). We prepare $|1\rangle$, perform measurements immediately followed by reset and another round of measurements. In the case of no errors we expect the second measurement to yield $|0\rangle$. We also perform the same sequence but without the first round of measurements and subtract that to remove the intrinsic reset and measurement errors. 
The results are shown in \cref{fig:results}(c). Complete ex-situ optimization adds on average an additional 0.4\% of reset errors, compared to 4.7\% and 1.4\% for partial and in-situ, respectively.

For the final benchmark, we quantify qubit leakage. Again, we focus on the measure qubits and prepare $|1\rangle$ as that makes the qubits more likely to leak, and then perform a variable number of measurements. We append a final and different measurement that is able to discriminate if the qubit has left the computational subspace. \Cref{fig:results}(d) shows the probability of leakage as a function of the number of measurements.
Complete ex-situ optimization suppresses leakage down to an average of 0.8\% after 20 measurement rounds, comparable to the heating limit as measured by repeating the experiment with no readout pulses but an equivalent amount of waiting time. After the same number of rounds, the partial and in-situ strategies have leakage populations of 5.7\% and 2.2\%, respectively.

Comparing the optimization strategies, we see that complete ex-situ using both predictive and heuristic models outperforms the others in all three benchmarks. It is able to achieve lower measurement errors, while also adding less reset and leakage errors for mid-circuit measurements. The partial ex-situ optimizer generally performs worse than in-situ optimization. This is likely due to the lack of an amplitude limiting model, which tends to drive the optimizer towards short and high-amplitude pulses, which in turn leads to state transitions. This emphasizes that for model-based optimization to work well, the models have to account for all dominant error mechanisms, even if only as heuristics.

In conclusion, we demonstrated model-based optimization for superconducting qubit readout achieving low measurement errors (1.5\%) for both mid-circuit and terminal measurements. For mid-circuit measurements, we also observed suppressed reset errors (0.4\%) and no increase in leakage due to readout. We accomplished this by overcoming the challenges stated in the introduction: the presented models accurately capture the relevant error channels, and they can be evaluated 10,000 times faster (1 minute vs 1 week for the parameter space used here) than measuring errors directly in hardware, which unlocks the ability to use a global optimizer. Based on recent work in Ref.~\cite{klimov2023snake} we believe the snake optimizer and these models will scale to at least 1,000 qubits.

Our model-based readout optimization strategy has already been employed in several large experiments, such as the demonstration of a distance-5 surface code \cite{google2023suppressing} with a measurement error of 1.9\% per qubit, and a 70 qubit random-circuit sampling experiment \cite{morvan2023phase} with an error of 1.3\% per qubit.
While the performance is among the best observed for repetitive and simultaneous measurements in superconducting qubits, even better performance will be needed to be well below the error-correcting threshold. In particular, the readout time has to be shorter to avoid data-qubit idling errors. 

We believe that the error rates achieved in this Letter are close to optimal for the given processor, and that the path to more performant readout is through longer relaxation times, higher measurement efficiencies, and more optimized circuit parameters. While we treated the circuit parameters as fixed, we could include them as optimization parameters to inform the design of future processors. However, more research is needed to find the optimal readout circuit for superconducting qubits.

\nocite{heinsoo2018rapid}

\begin{acknowledgments}
We thank the broader Google Quantum AI team for fabricating the processor, building and maintaining the cryogenic system, and general hardware and software infrastructure that enabled this experiment.  We also thank V.~Sivak and W.~Livingston for providing comments on the manuscript, and L.~Martin for helping with the crosstalk analysis.
\end{acknowledgments}

\bibliography{bib}
\end{document}


\title{Supplemental materials: Model-based Optimization of Superconducting Qubit Readout}
\author{Andreas Bengtsson}
\author{Alex Opremcak}
\author{Mostafa Khezri}
\author{Daniel Sank}
\author{Alexandre Bourassa}
\author{Kevin J. Satzinger}
\author{Sabrina Hong}
\author{Catherine Erickson}
\author{Brian J. Lester}
\author{Kevin C. Miao}
\affiliation{Google Quantum AI, Santa Barbara, CA}

\author{Alexander N. Korotkov}
\affiliation{Google Quantum AI, Santa Barbara, CA}
\affiliation{Department of Electrical and Computer Engineering, University of California, Riverside, CA}

\author{Julian Kelly}
\author{Zijun Chen}
\author{Paul V. Klimov}
\affiliation{Google Quantum AI, Santa Barbara, CA}

\date{\today}

\maketitle

\clearpage

\section{Measurement setup}
Overall, the control and measurement setup of the processor is similar to as described in \cite{arute2019quantum}.
On a 72 qubit Sycamore processor there are in total 12 readout lines, which run diagonally across the device. 
Each readout line has a Purcell filter, which is shared among 6 qubits; however, since in this work we only use 17 qubits in a distance-3 surface code arrangement we maximally use 3 qubits on a single readout line.
Furthermore, each readout line has a near quantum-limited parametric amplifier operating in the phase-preserving mode. The frequencies of the resonators (see \cref{fig:fixed_params}) are chosen with respect to the bandwidths of the filter, the amplifier, and the electronics used to synthesize and sample the readout signals. In the same figure the readout lines are indicated by dashed diagonal lines.

\section{Optimized parameters}
\begin{figure}
    \centering
    \includegraphics[width=0.75\linewidth]{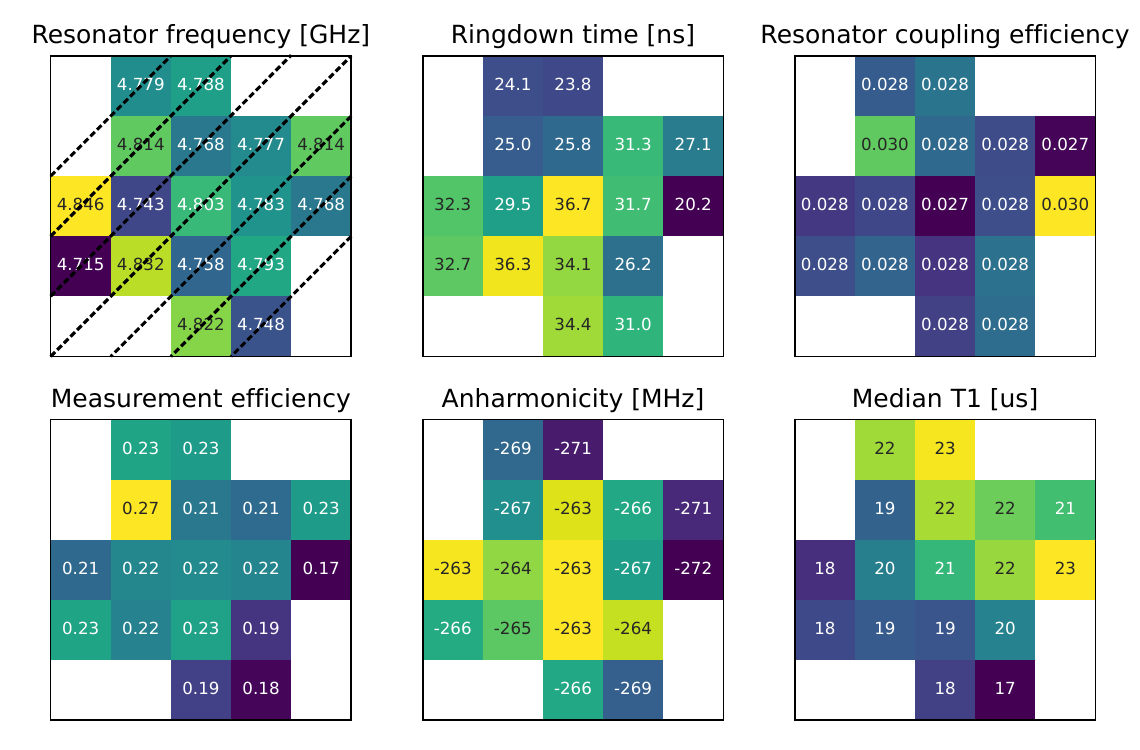}
    \caption{Individual qubit values for the fixed parameters. The color of each square represents the value for that qubit, but note that the color scale is not shared between panels. In the first panel, the individual readout lines are indicated by dashed diagonal lines. Note that for the qubit-resonator coupling we plot the frequency-independent coupling efficiency, which when multiplied by $\sqrt{\omega_r\omega_q}/2$ gives the coupling strength $g$. For $T_1(\omega_q)$ we show the median value across the frequency range used in the optimization.}
    \label{fig:fixed_params}
\end{figure}

\begin{figure}
    \centering
    \includegraphics[width=1\linewidth]{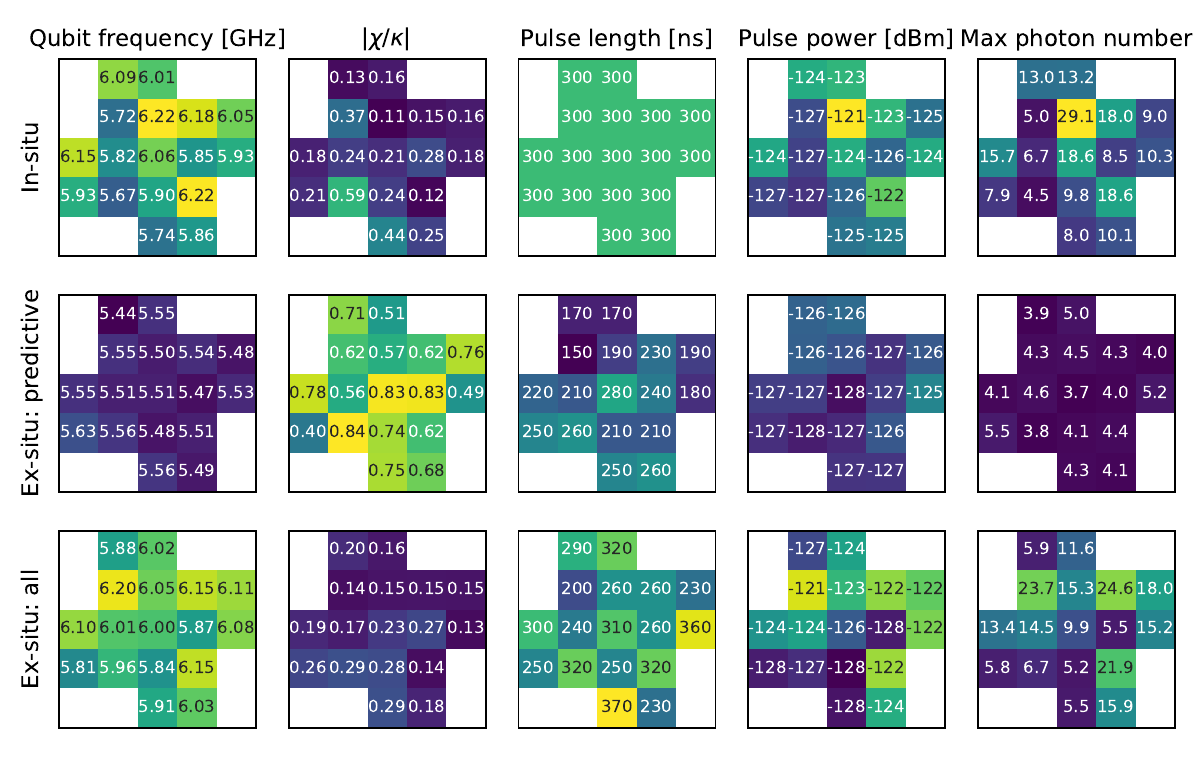}
    \caption{
    Parameter outputs of the three optimization strategies (rows). The pulse power is referred to the input of the readout resonators, and the maximum photon number in the resonator is calculated using Eq.~(2) in the main text. Color scales are shared within each column.
    }
    \label{fig:opt_params}
\end{figure}

The fixed parameters which the error models take as input can be seen in \cref{fig:fixed_params}.
For completeness we plot the parameter outputs of the three optimization strategies in \cref{fig:opt_params}, but it is not easy to draw many quantitative conclusions from the values. One interesting thing to note in the ex-situ optimization with only predictive models (middle row) is that the qubits are very close in frequency, especially along the diagonals. This is not ideal since there is non-negligible coupling between the qubits (as described in the main text) and thus they can be subject to swapping, which leads to measurement errors.
When we include the heuristic qubit-qubit coupling model (bottom row) we see that the frequencies are more spread out.

Another quantity of interest is the ratio between the dispersive shift, $\chi$, and the resonator linewidth, $\kappa$. From an SNR perspective and given a fixed readout amplitude, the optimal value is $\chi/\kappa=0.5$. We expect that ex-situ optimization using only predictive models would yield qubit frequencies that are close to achieving that ratio, which we see to a large extent. When using all models, we see that $\chi$ is typically smaller, which is due to the fact that the heuristic measurement-induced state transition models tend to push qubits to larger frequencies where $\chi$ is smaller.

A similar analysis can be done regarding the maximum photon number. Using predictive only models we see that the maximum photon numbers are quite uniform across all 17 qubits, while for all models they have a larger spread and typically reach higher photon numbers as well. This is again due to the measurement-induced state transitions which forces the qubits to use larger frequencies where more photons are required to achieve the same SNR. The inclusion of the crosstalk model also forces a spread in qubit frequencies, which in turn causes a spread in $\chi$ and maximum photon numbers.

\section{The snake optimizer}
\begin{figure}
    \centering
    \includegraphics[width=1\linewidth]{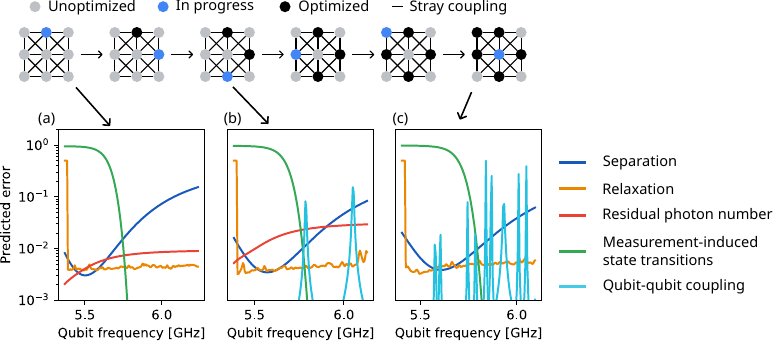}
    \caption{Graph traversal and build-up of the cost functions for 9 qubits with nearest and next-nearest neighbor couplings. We start with a fully unoptimized processor. We choose a starting qubit and optimize it, then traverse to a new one while adding relevant error models as we go, until we optimized all qubits. 
    (a) Error models for the first (measure) qubit. The measurement-induced state transition is implemented as a smoothed step function when the maximum photon number goes above a certain value (see the main text for more details).
    (b) Error models for the second (measure) qubit. Here, error components for coupling to the first qubit has been added.
    (c) Error models for a later (data) qubit with a dense spectrum of error components. Note the absence of the reset error model since this qubit doesn't have mid-circuit measurements.
    }
    \label{fig:optimization}
\end{figure}

Here, we briefly describe the workflow of the snake optimizer as shown in \cref{fig:optimization}. We start with all qubits unoptimized. Then, we pick a start qubit, construct its cost function [\cref{fig:optimization} (a)], and find the global minimum using a brute-force search over the three parameters. Those parameters are now locked in for that qubit. We then traverse to a second qubit and repeat the process, with the addition of the qubit-qubit coupling models, which depend on the parameters of the first qubit. As seen in \cref{fig:optimization} (b) and (c), the frequency spectrum gets more crowded as the process goes on. This means that as we traverse the processor, the more constraint the optimization problem becomes; however the constraints do not grow indefinitely since we only include up to next-nearest neighbors. For the surface code, where low readout error are more important for the measure qubits, we start on a measure qubit and traverse diagonally (to another measure qubit); this allows those qubits to explore a larger search space and find a solution with lower error. When all measure qubits have been optimized, we optimize the data qubit until the full processor is completed.

\section{Detailed benchmarking results}
\begin{figure}
    \centering
    \includegraphics[width=0.8\linewidth]{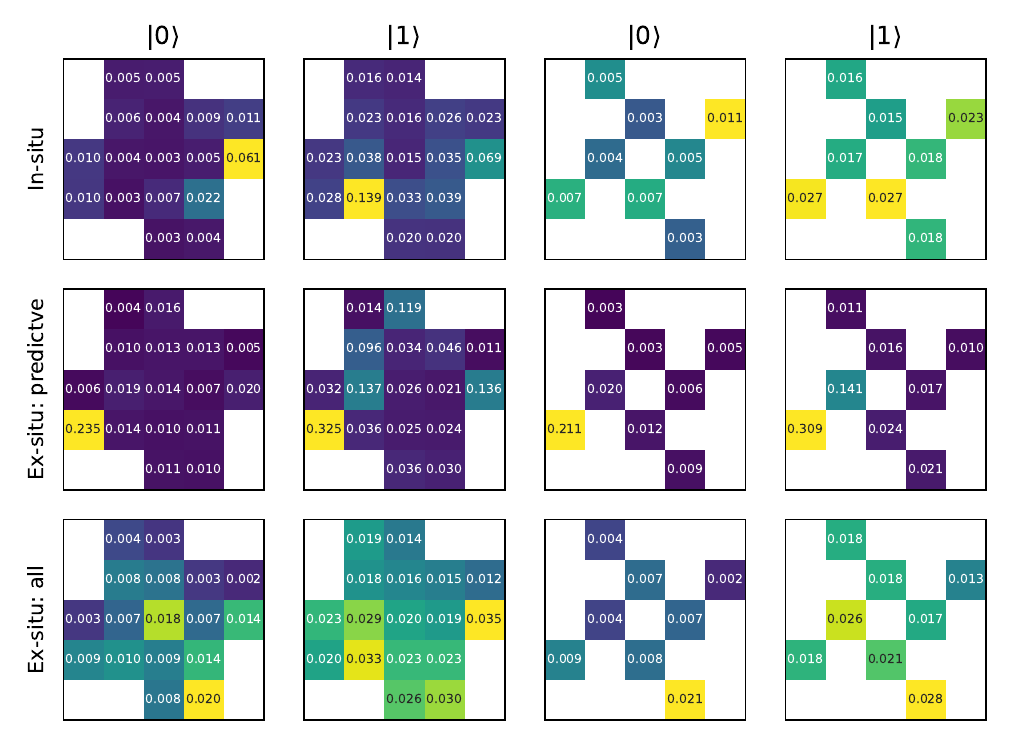}
    \caption{
    Measurement errors for each qubit. The rows corresponds to different optimization strategies, and the first two columns are the errors for $|0\rangle$ and $|1\rangle$, respectively. The last two columns are also measurement errors for the same states, but in the case when only the measure qubits are measured.
    Note that the color scale is not shared between panels.}
    \label{fig:measure_errors}
\end{figure}
\begin{figure}
    \centering
    \includegraphics[width=0.6\linewidth]{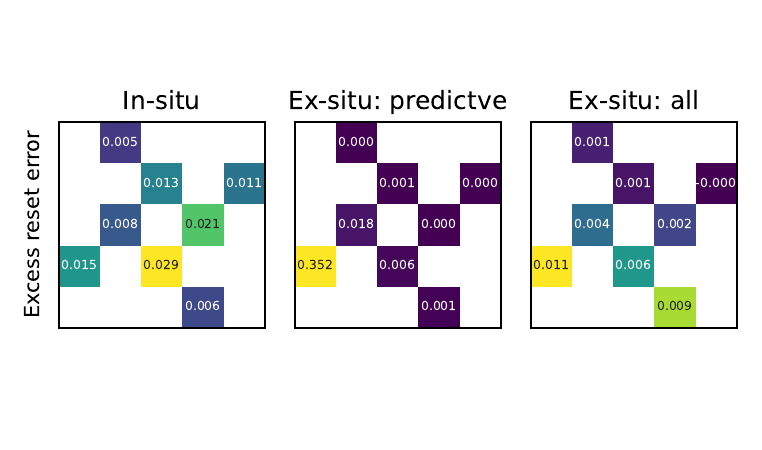}
    \caption{Excess reset error for each measure qubit and optimization strategy.
    Note that the color scale is not shared between panels.
    }
    \label{fig:reset_errors}
\end{figure}
\begin{figure}
    \centering
    \includegraphics[width=1\linewidth]{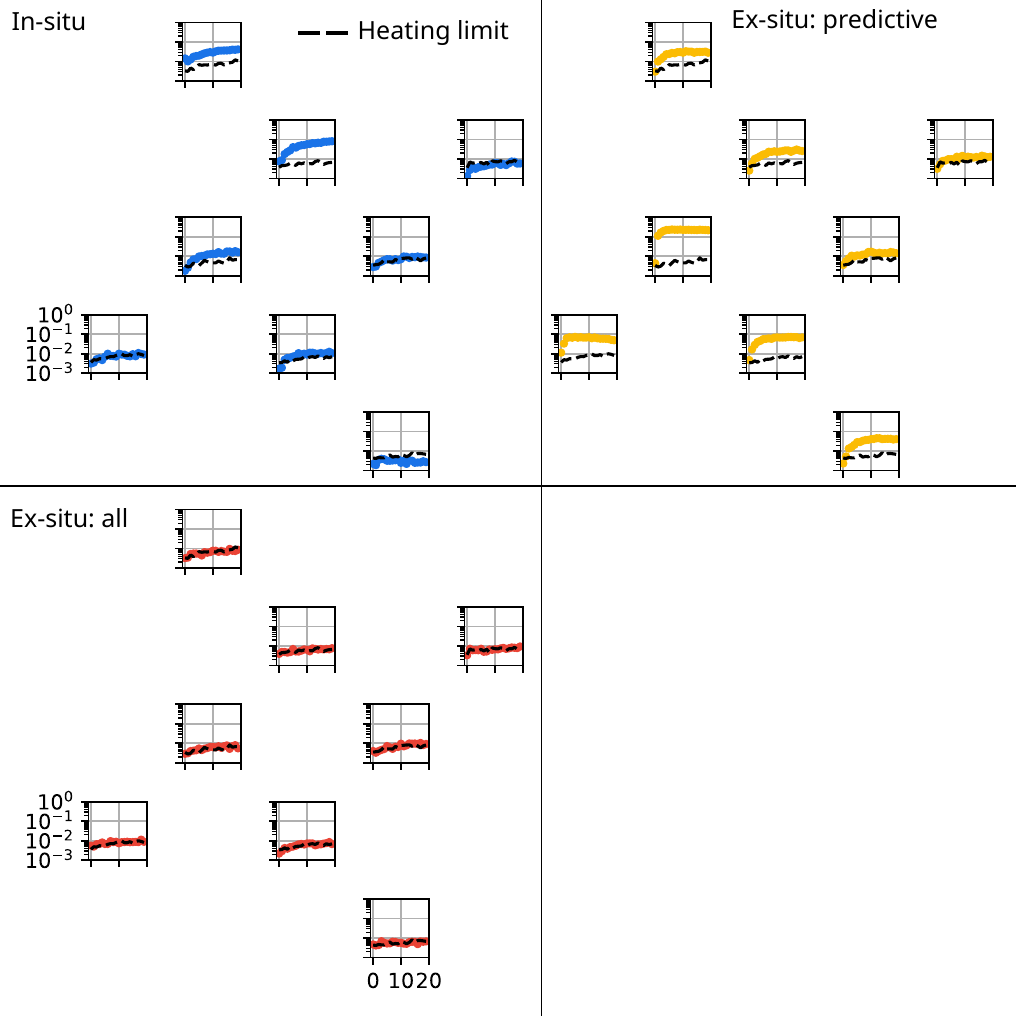}
    \caption{Probability of qubit leakage after $N$ rounds of measurements for each measure qubit and optimization strategy. All panels share the same x and y axes.}
    \label{fig:leakage}
\end{figure}

In \cref{fig:measure_errors,fig:reset_errors,fig:leakage} we show the benchmarking results on a per qubit basis. In \cref{fig:measure_errors} for each optimization strategy we separate the errors for $|0\rangle$ and $|1\rangle$, as well as for the all qubits simultaneously and measure qubits only cases. For instance, we can see the large outlier of 0.139 for in-situ $|1\rangle$, which is due to a frequency collision between that qubit's $|0\rangle \leftrightarrow |1\rangle$ transition and the qubit to its right's $|1\rangle \leftrightarrow |2\rangle$ transition, as can be seen in \cref{fig:opt_params}. 

Another large outlier is the bottom left qubit in the ex-situ optimization with predictive models only. By looking in \cref{fig:reset_errors} we see that the same qubit is a large outlier in reset error as well, indicating that this is due to severe measurement-induced state transitions. We can confirm this by inspecting \cref{fig:leakage} where the same qubit has a large probability of being leaked after just one measurement.

\section{Crosstalk}
\begin{figure}
    \centering
    \includegraphics[width=0.5\linewidth]{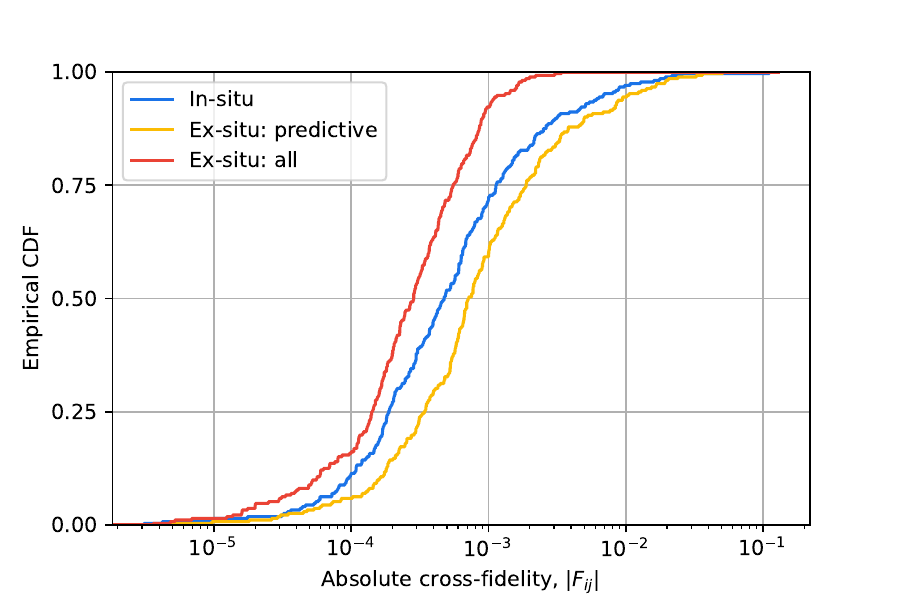}
    \caption{Integrated histograms of the absolute values of the off-diagonal cross-fidelity matrix elements, $|F_{ij}|$, for the three optimization strategies.
    }
    \label{fig:crosstalk}
\end{figure}

In addition to just looking at the measurement error for each qubit and try to infer the impact of crosstalk we can directly look at correlations between qubits. A variety of different metrics exist, with one being the cross-fidelity matrix \cite{heinsoo2018rapid}. In this work have to use a slightly modified formula which properly accounts for the fact that we are only sampling $200$ out of the $2^{17}=131,072$ possible states,
\begin{equation}
    F_{ij} = 1 - \left[ P(1_i | 0_j) + P(0_i | 1_j) \right] = 1 - \frac{1}{2} \left[ P(1_i | 0_i 0_j) + P(1_i | 1_i 0_j) + P(0_i | 1_i 1_j) + P(0_i | 0_i 1_j) \right],
\end{equation}
where $P(x_i | y_i z_j)$ is the probability to measure state $x$ for qubit $i$ given that it was prepared in state $y$ while qubit $j$ was prepared in state $z$. For an ideal measurement system ${F_{ij} = 0 \quad \forall \quad i \ne j}$. In \cref{fig:crosstalk} we plot the distributions of $|F_{ij}|$ for all off-diagonal elements and the three optimization strategies. As expected, ex-situ using all available error models is outperforming the other two, achieving a mean of 0.04\%, compared to 0.19\% and 0.24\% for in-situ and ex-situ: predictive only, respectively. This is expected since the crosstalk error model described in the main text is only included in the ex-situ: all optimization, which means that there is nothing preventing qubit frequency collisions in the other two strategies.

\section{Error budget}
\begin{figure}
    \centering
    \includegraphics[width=0.4\linewidth]{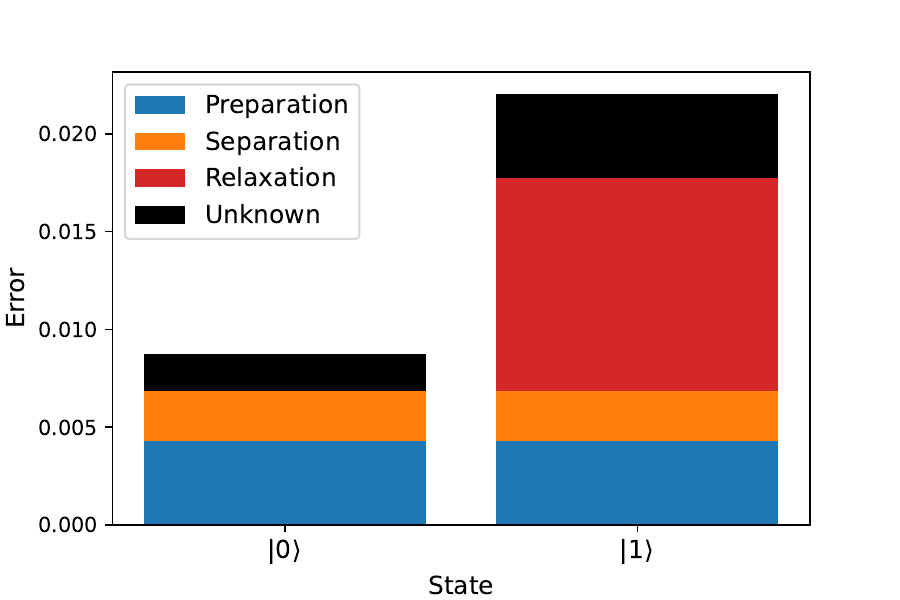}
    \caption{Error budgets for the $|0\rangle$ and $|1\rangle$ states averaged over all 17 qubits. Each color corresponds to a different error mechanism. The differences between the actually observed error and the three considered mechanisms are indicated by ``unknown''.
    }
    \label{fig:error_budget}
\end{figure}

To provide insight for what contributes to the observed 1.5\% error per qubit (for the ex-situ optimization with all models) we construct an error budget (seen in \cref{fig:error_budget}). We consider three possible error mechanisms:
\begin{itemize}
    \item Separation error [see Eq.~(4) in the main text]. We estimate this from the separation and distribution widths in phase-space when preparing $|0\rangle$ and $|1\rangle$ on every qubit
    \item State preparation error. We take this as the probability that we did not prepare $|0\rangle$. We estimate it by preparing $|0\rangle$, measure the probability of $|1\rangle$, and then subtract the separation error. Ideally we would also include the error of preparing $|1\rangle$, however this is more difficult to faithfully extract so we choose to not include it here. For reference, the single-qubit error as measured by randomized benchmarking earlier on the same device was close to 0.001 \cite{google2023suppressing}.
    \item Relaxation error. We calculate this using Eq.~(6) in the main text.
\end{itemize}
From these three error mechanisms we estimate an error of 1.2\% per qubit. The remaining error is indicated in \cref{fig:error_budget} as unknown and could include errors from sources like crosstalk, preparation of $|1\rangle$, and measurement-induced state transitions. With the tools currently available to us we are not able to quantify the contributions from these separately.

\clearpage
\bibliography{bib}